\documentclass[prl,twocolumn,tightenlines,floatfix,nofootinbib]{revtex4}
\usepackage{graphicx,amsfonts,amssymb,amsmath}

\def\be{\begin{equation}}
\def\ee{\end{equation}}
\def\ba{\begin{eqnarray}}
\def\ea{\end{eqnarray}}
\def\ge{\mathrel{\raise.3ex\hbox{$>$\kern-.75em\lower1ex\hbox{$\sim$}}}}
\def\la{\mathrel{\raise.3ex\hbox{$<$\kern-.75em\lower1ex\hbox{$\sim$}}}}

\def\simgt{\mathrel{\raise.3ex\hbox{$>$\kern-.75em\lower1ex\hbox{$\sim$}}}}
\def\simlt{\mathrel{\raise.3ex\hbox{$<$\kern-.75em\lower1ex\hbox{$\sim$}}}}

\newcommand{\bi}[1]{\bibitem{#1}}
\newcommand{\fr}[2]{\frac{#1}{#2}}

\newcommand{\nc}{\newcommand}

\nc{\gone}{\bar g_{\pi NN}^{(1)}}
\nc{\gzero}{\bar g_{\pi NN}^{(0)}}
\nc{\al}{\alpha}
\nc{\ga}{\gamma}
\nc{\de}{\delta}
\nc{\ep}{\epsilon}
\nc{\ze}{\zeta}
\nc{\et}{\eta}
\nc{\ka}{\kappa}
\nc{\rh}{\rho}
\nc{\si}{\sigma}
\nc{\ta}{\tau}
\nc{\up}{\upsilon}
\nc{\ph}{\phi}
\nc{\ch}{\chi}
\nc{\ps}{\psi}
\nc{\om}{\omega}
\nc{\Ga}{\Gamma}
\nc{\De}{\Delta}
\nc{\La}{\Lambda}
\nc{\Si}{\Sigma}
\nc{\Up}{\Upsilon}
\nc{\Ph}{\Phi}
\nc{\Ps}{\Psi}
\nc{\Om}{\Omega}
\nc{\ptl}{\partial}
\nc{\del}{\nabla}
\nc{\ov}{\overline}
\nc{\newcaption}[1]{\centerline{\parbox{15cm}{\caption{#1}}}}

\begin{document}

\title{Pseudoscalar perturbations and polarization of the cosmic microwave background
}
\author{Maxim Pospelov$^{1,2}$, Adam Ritz$^{1}$ and Constantinos Skordis$^{2}$}
\affiliation{$^1$Department of Physics and Astronomy, University of Victoria,
             Victoria, BC, V8P 5C2 Canada \\
 $^{2}$Perimeter Institute for Theoretical Physics, Waterloo,
ON, N2L 2Y5, Canada}
\date{August 2008}
\begin{abstract}
\noindent
We show that models of new particle physics containing massless pseudoscalar fields 
super-weakly coupled to photons can be very efficiently probed with CMB polarization
anisotropies. The stochastic pseudoscalar fluctuations generated during inflation provide a 
mechanism for converting $E$-mode polarization to $B$-mode during photon propagation
 from the surface of last scattering. The efficiency of this conversion process is controlled by 
 the dimensionless ratio $H/(2\pi f_a)$, where $H$ is the Hubble scale during inflation, and $f_a^{-1}$ is the strength 
of the pseudoscalar coupling to photons. The current observational limits on the $B$-mode 
constrain this ratio to be less than 0.07, which in many models of inflation 
translates to a sensitivity to values of  $f_a$ in excess of $10^{14}$ GeV, surpassing the 
sensitivity of other tests.

\end{abstract}
\maketitle

{\it Introduction}:--- Within the last decade, precision observations of 
fluctuations in the cosmic microwave background (CMB)
have provided us with a powerful probe of cosmology, 
and been the driving force in pinning down the parameters in the
concordance $\La$CDM model. In recent times, gains in sensitivity 
have allowed us to complement observations of the temperature fluctuations 
 with those in various components of polarization. 
Much of the focus has been aimed at the detection of the so-called
$B$-mode of polarization, as it is an important probe of 
primordial gravitational waves, and thus of the possible mechanism for
inflation. However, precision probes of polarization in various modes 
also present us with a powerful set of tools to test for new degrees of 
freedom with which the CMB photons may interact.  In this Letter, we show 
that existing upper bounds on the $B$-mode of polarization already place 
stringent constraints on any new light pseudoscalar degrees of freedom
through their stochastic fluctuations generated during inflation.

The interactions of a massless (or nearly massless) 
pseudoscalar field $a$ with photons may be parametrized in terms of  a 
dimensionful coupling constant $f_a^{-1}$,
\be
\label{Lagr}
 {\cal L}_{\gamma a} = -\frac{1}{4} F_{\mu\nu} {F}^{\mu\nu}
+\frac{1}{2}\ptl_\mu a\ptl^\mu a + \frac{a}{2f_a} F_{\mu\nu} \tilde{F}^{\mu\nu}
\ee
where $F_{\mu\nu}$ is the electromagnetic field strength tensor, and $\tilde{F}_{\mu\nu}$ its dual. 
We note that the absence of a potential, $V(a)=0$, is protected by the shift symmetry of this 
Lagrangian, $a \to a +{\rm const}$, and while it is also natural to expect that the 
pseudoscalar $a$ would have derivative couplings to the
spins of other particles,  this does not affect the physics of the CMB. 
The presence of the photon coupling implies that a space-time variation of the field 
$a$  will induce a rotation $\De \psi$ of 
the polarization of an electromagnetic wave. Provided the photon wavelength is much smaller than the 
characteristic scale on which the field $a$ varies \cite{hs}, the rotation angle is 
independent of frequency: 
\be
\label{extrarot}
\De \psi = \frac{1}{f_a}\De a.
\ee
In the past, CMB polarization has been used to probe an induced
rotation angle due to a spatially homogeneous pseudoscalar profile $a=a(t)$ \cite{hs,lwk}, 
with the latest data \cite{cmb_rot2,wmap5} giving comparable sensitivity to  
the strongest constraints from 
polarized extra-galactic radio sources \cite{cfj}. However, the time evolution of the 
background effectively amounts to a bulk violation of parity and/or Lorentz invariance. Here, we
 consider a scenario perfectly in keeping with standard cosmology, and 
focus on a stochastic background for the pseudoscalar as
would arise from a period of inflation.  We 
exploit the fact that all massless scalars  have a universal amplitude for fluctuations generated
during inflation, $\de a = H/(2\pi)$, where $H$ is the inflationary Hubble parameter
at horizon crossing. Without loss of generality, we are allowed 
to choose $a_{\rm now} = 0$, and then the additional rotation $\psi$ (\ref{extrarot}) 
for a photon arriving from the direction $\hat n$ 
is simply given by the randomly fluctuating value of $A(\hat n, \tau) \equiv
a(\hat n, \tau)/f_a$ at the surface of last scattering (LSS), $\tau \simeq \tau_{\rm LSS}$,
where $\ta$ is the conformal time. 
To quantify the effect, 
we introduce the dimensionless parameter $c_a$:
\be
c_a = \left(\frac{H}{2\pi f_a}\right)^2 , ~~|\De\psi| \sim \sqrt{c_a} . \label{a_norm}
\ee
It is then apparent that any ${\cal O}(1)$ probe of these fluctuations will provide impressive 
sensitivity to $f_a$ on the order of the inflationary Hubble scale. For a scale-invariant source of
inflationary perturbations in the field $a$, the evolution of the rotation angle $\psi$ with
the scale factor $R(t)$ can be pictured as an
effective random walk where each efold of expansion corresponds to a separate step,
so that the cumulative effect can be further enhanced relative to (\ref{a_norm}): 
$(\De \ps)^2 \sim c_a \ln(R_{\rm now}/R)$. 

In what follows, we analyze the generation of the $B$-mode of CMB polarization from the inflationary fluctuations of  a massless pseudoscalar. We use the
$B$-mode spectrum to deduce a precise limit on $c_a$,  and finally describe its implications for models of particle physics incorporating
new pseudoscalar fields. We should note that the impact of inflationary axion perturbations has previously been discussed in the context of
isocurvature perturbations on the CMB temperature spectrum \cite{tw}, but this is not directly relevant for non-axionic massless pseudoscalars.  
The technical aspects of  our analysis do nonetheless have parallels with  studies of 
the impact on the CMB spectra  of stochastic magnetic fields \cite{mag}, of scalar moduli \cite{skk},
and of weak lensing, in that the convolution of an additional fluctuating 
source with the underlying density perturbations tends to mix the various
power spectra for temperature and $E$ and $B$-mode polarization. 

{\it Generating $B$-mode polarization}:---At the LSS 
the underlying spectrum of scalar (and tensor) perturbations leaves its imprint on 
the CMB through anisotropies in temperature and linear polarization, 
the latter generated by Thomson scattering.  Linear polarization is 
conveniently described in terms of the $Q$ and $U$ Stokes
parameters, and while a single scalar perturbation mode
with $\vec{k}||\hat z$ does not generate $U$, it generates $Q$. 
We  generalize the approach of Zaldarriaga and Seljak \cite{zs} where,
for a mode of momentum $k$ directed along $\hat z$, the observed value of 
$Q$ in a given direction $\hat{n}$  can be written as an integral over 
conformal time $\ta$ along the line of sight,
\be
 Q(k,\hat{n}) = \frac{3}{4}(1-\mu^2)\int_0^{\tau_0} e^{ix\mu} g(\ta)\Pi(k,\ta).
\ee
Here  $x=k(\ta_0-\ta)$ and $\mu=\hat{k}\cdot\hat{n}$. The source and visibility 
functions, $\Pi(k,\ta)$ and $g(\ta)$ \cite{zs}, reflect the details of 
the generation of polarization via Thomson scattering at the LSS. 
In the presence of a small stochastic rotation (\ref{extrarot}),
a $U$ component would also be generated, $U\simeq 2\De\psi Q = 2AQ$. This rotation also depletes 
$Q$, but this is a less significant effect.
It then follows that for the combination of a single Fourier mode of the underlying scalar perturbation, 
with $\vec{k}||\hat z$, and one Fourier mode of the pseudoscalar perturbation with arbitrary direction $\vec{q}$,
the result for  $U$ can be exressed as 
\be
U(k, q, \hat{n}) = \frac{3}{2}(1-\mu^2)\int_0^{\tau_0} 
e^{ix\mu+iy\nu} g(\ta) \Pi(k,\ta)\De_A(\ta,q).
\ee
In this expression, $y=q(\ta_0-\ta)$ and $\nu=\hat{q}\cdot\hat{n}$.
 $\De_A({q},\ta)$ is the transfer function, that
has a generic form for any massless (pseudo)scalar field:  $\De_A({q},\ta)$ is normalised to one on large scales,
and oscillates with a decaying envelope determined by the matter content on sub-horizon scales. 
At the next step, we calculate basis-independent expansion coefficients by performing the 
projection onto spin 2-weighted spherical harmonics,
$a_{Blm}=-\int d\Om (Y_{2,lm}^* + Y_{-2,lm}^*)U(\hat{n})/2$,
as described in \cite{zs},
\ba
 a_{Blm} &=& \frac{3}{2}\left[\fr{(l-2)!}{(l+2)!}\right]^{1/2} \int d\Om_n d^3k d^3q Y_{0,lm}^*(\hat{n}) 
\nonumber\\
  && \!\!\!\!\!\!\!\!\!\!\!\!\!\!\!\!\!\! \times \int_0^{\ta_0} 
d\tau (m^2-(1+\ptl_x^2)^2 x^2) e^{ix\mu+iy\nu} F(\ta,\vec{k},\vec{q}).
\label{ablm}
\ea
Since the $a_{Blm}$'s are basis-independent, we are now allowed to generalize the scalar-pseudoscalar 
source function to modes of arbitrary direction $\vec{k}$ and $\vec{q}$, which are in turn determined by the 
{\em primordial} scalar $\xi(\vec{k})$ and pseudoscalar $\xi_A(\vec{q})$
perturbations:
\be 
 F(\ta,\vec{k},\vec{q})= g(\ta)\Pi(k,\ta)\De_A(q,\ta) \xi(\vec{k})\xi_A(\vec{q}).
\ee
The standard assumption that these perturbations are gaussian random variables implies 
\ba
\langle \xi^*(\vec{k}_1) \xi(\vec{k}_2)\rangle &=& P_\phi(k_1) \de^{(3)}(\vec{k}_1-\vec{k}_2),\nonumber \\
\langle \xi_A^*(\vec{q}_1) \xi_A(\vec{q}_2)\rangle &=& P_A(q_1) \de^{(3)}(\vec{q}_1-\vec{q}_2),
\ea
where $P_\phi$ is the (now rather well-measured) primordial power spectrum for scalar perturbations. The scalar 
primordial power spectrum $P_\phi$  sources 
the CMB anisotropy and indeed structure formation, while the pseudoscalar power spectrum 
$P_A$ is, in our approach,  directly linked to $c_a$ in (\ref{a_norm}) by the standard inflationary prediction,
\be
P_A(q) = \fr{c_a}{4\pi q^{3}} q^{n_a-1}.
\label{P_A} 
\ee
Here $n_a$ is the spectral index of the pseudoscalar primordial power spectrum, with $n_a=1$ being scale invariant. 
Note that the factor of $4\pi$ reflects  the normalization convention for Fourier transforms in \cite{zs} that we follow in this paper. 

To proceed, we expand the exponential $e^{ix\mu+iy\nu}$ in (\ref{ablm})
into a basis of products of spherical harmonics, and using their orthogonality and 
completeness, perform all the angular
integrals.
Deferring the remaining details \cite{prs2}, we
obtain
\ba
 C_{Bl} &=& \frac{1}{2l+1}\sum_m \langle a_{Blm}^* a_{Blm}\rangle = \frac{4(4\pi)^3}{2l+1} \frac{(l-2)!}{(l+2)!} \nonumber\\
&&  \times \sum_{m,l_1,l_2} (2l_1+1)(2l_2+1)\left(\begin{array}{ccc} l & l_1 & l_2 \\ 0 & 0 & 0 \end{array}\right)^2 \nonumber \\
 && \;\;\;\; \times \int k^2 P_\Ph  q^2 P_A dkdq |\De_{l_1l_2m}(k,q)|^2,
\ea
with the generalized transfer function,
\ba
\label{gtf}
 \De_{l_1l_2m}(k,q) &=& \frac{3}{4}\int_0^{\ta_0} d\ta g(\ta)  j_{l_1}(x)j_{l_2}(y) \nonumber\\
  && \!\!\!\!\!\!\!\!\!\!\!\!\!\!\!\!\!\!\!\!\! \times\left(\frac{(l_1+2)!}{(l_1-2)!}\frac{1}{x^2} - m^2\right) \De_A(\ta,q) \Pi(\ta,k).
\ea
The sum over $m$ can be straightforwardly computed, while the 
summation range for $l_1$ and $l_2$ is restricted to $|l_1-l_2|\leq l \leq l_1+l_2$,
which is enforced by the Wigner 3$j$-symbol. It is important to note that the cross-correlations
$TB$ and $EB$ vanish on account of the overall conservation of parity.

\begin{figure}[t]
\centerline{\includegraphics[width=8.7cm]{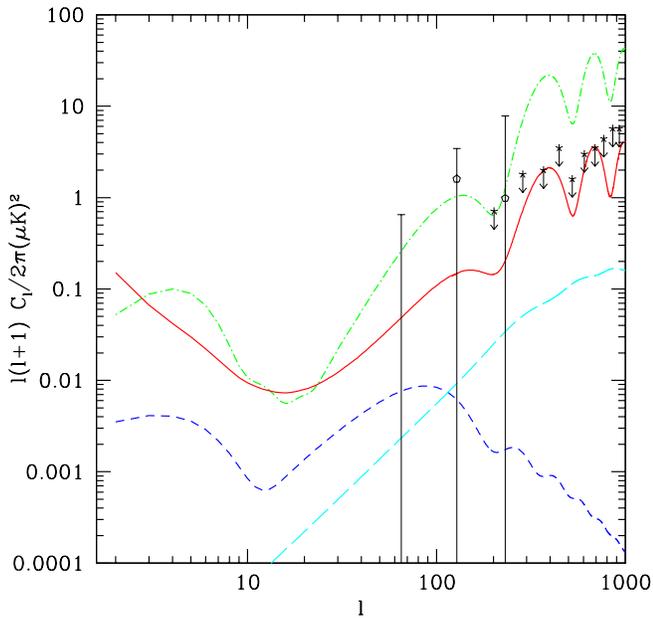}}
\caption{We display the $C_l$'s for the $E$-mode (green -- dot-dash) and induced $B$-mode polarization (red -- solid). 
The fiducial value $c_a=4.2\times 10^{-3}$ was
chosen to match the current upper bound on the $B$-mode using 
data from QUaD. 95\% upper limits from QUaD (stars) and $2\sigma$ upper limits from WMAP5 (polygons) are shown for 
comparison. We also exhibit the $C_{Bl}$'s induced by lensing 
of the $E$-mode (cyan -- long-dash), and the contribution from primordial gravitational waves with $r=0.14$ (blue -- dash),
corresponding to the choice $H_{14} = 1$.}
\label{fig1}
\end{figure}

At this point we assume a  
spectrum of scalar fluctuations with scalar spectral index $n=0.963$ chosen according to the best fit 
model for the WMAP5 CMB data, and take the pseudoscalar spectral index $n_a$ to have  the same value. 
We used our own code based on {\tt CMBfast} \cite{sz} to numerically calculate the functions $\Pi(k,\tau)$ and $\Delta_A(\tau,q)$
for the WMAP5 best fit model,
and then to compute $C_{Bl}$  using Eqs.~(\ref{P_A})--(\ref{gtf}).
Fig.~\ref{fig1} displays the results for $C_{Bl}$ for a fiducial choice 
of $c_a=4.2\times 10^{-3}$. For comparison, the plot also shows $C_{El}$ as well as the 
$B$-modes generated by primordial tensor perturbations with $r=0.14$ and
by lensing of the $E$-mode. The qualitative form of $C_{Bl}$ induced by 
pseudoscalar perturbations can be understood by looking at the dominant regions in the 
$l_1$ and $l_2$ sums. For large $l$, we find that $l_1\sim l$, while the sum over 
$l_2$ is effectively truncated at a lower value of
order $l_2^{\rm max} \sim (\ta_{\rm 0}- \ta_{\rm LSS})/\ta_{\rm LSS} \sim 50$ (higher values of $l_2$ contribute 
no more than about 2\% to the $C_{Bl}$'s on small scales). 
Thus for large $\l \sim 1000$ the induced $B$-mode closely tracks the 
underlying $E$-mode. For lower values of $l\sim {\cal O}(1)$, both 
$l_1$ and $l_2$ saturate at higher scales so the $B$-mode is somewhat
larger. Finally, the overall scale of the oscillations in $C_{Bl}$ is 
slightly suppressed as is to be expected from convoluting the underlying
$E$-mode source with a gaussian random variable. 

It is important to keep in mind that our approximation, which assumes a small rotation 
angle, may break down once $C_{Bl}$ becomes comparable to  $C_{El}$. Consequently, for setting constraints
on $C_{Bl}$ we choose $l$ in the interval 100--1000, where the most recent 
experimental results of QUaD \cite{quad} 
probe the $B$-modes well below the detected $E$-mode level, $C_{Bl} \sim {\cal O}(0.1)\times C_{El}$.  
The QUaD limits, shown in Fig.~\ref{fig1}, impose a stringent constraint on $c_a$: 
\be
 c_a < 4.2\times 10^{-3} \;\;\;\; \Longrightarrow \;\;\;\; f_a > 2.4 \times 10^{14}~{\rm GeV}\times H_{14}, \label{const}
\ee
where we have introduced $H_{14} \equiv H/10^{14}$ GeV, a normalization 
inspired by the fact that  $H_{14} \sim O(1)$ is close to the maximal inflationary value 
of $H$ allowed by observations. The inflationary Hubble scale  
can be traded for the tensor-to-scalar ratio $r$, commonly used to parametrize the strength of 
all massless perturbations including gravitational waves, $r=0.14\times H_{14}^2$.  
Note that from Eq.~(\ref{a_norm}), the conclusion that 
$c_a \ll 1$ justifies a posteriori our perturbative treatment 
of the rotation of polarization. The constraint (\ref{const}) is the main result of this Letter, and in 
what follows we discuss its implications.

{\it Implications for particle physics}: Given the conventional picture of inflationary cosmology, 
the constraint obtained 
above applies to massless pseudoscalars  (or almost massless with a mass 
below the Hubble scale at decoupling) and it is important to
consider how such new low-energy degrees of freedom may 
naturally arise. Recall that massless pseudoscalar fields coupled to the 
operator $G_{\mu\nu} \tilde{G}^{\mu\nu}$ in QCD, namely axions,  resolve the strong $CP$ problem in a natural way \cite{ww}
while gaining an anomaly-induced mass $m_a \sim m_\pi f_\pi/f_a$. While this mass is large compared to the scales 
relevant here, this mechanism for inducing a mass is unique, and thus should two or more  pseudoscalars couple to 
the QCD anomaly, only one linear combination would become massive \cite{AU}. Schematically, 
below the QCD scale, such models lead to an effective Lagrangian of the form,
\ba
\left(\frac{a_1}{2g_{1}} + \frac{a_2}{2g_{2}}\right)G_{\mu\nu} \tilde{G}^{\mu\nu} +
\left(\frac{a_1}{2f_{1}} + \frac{a_2}{2f_{2}}\right)F_{\mu\nu} \tilde{F}^{\mu\nu} \nonumber 
\\\to {\cal L}_{{\rm QCD}a} + \frac{a}{2f_{a}}F_{\mu\nu} \tilde{F}^{\mu\nu},
\ea
where besides the Lagrangian ${\cal L}_{{\rm QCD}a}$ for the QCD-axion, one has a 
massless pseudoscalar $a$ as part of ${\cal L}_{\gamma a}$ in (\ref{Lagr}) with 
$f_a^{-1} = (g_2/f_2-g_1/f_1)/\sqrt{g_1^2+g_2^2}$. In the absence of any special reasons for 
cancellation, the generic case is $f_a^{-1} \neq 0$. Although we refrain from assessing the likelihood 
of new high-energy physics leading to two or more light pseudoscalar fields, we can refer to various
scenarios in string theory where multiple
pseudoscalar moduli are a generic prediction (see {\em e.g.} \cite{SW}). 
Therefore, the Lagrangian (\ref{Lagr}) can easily originate from a more fundamental theory with
multiple degrees of freedom.  Note that it would take a significant increase in sensitivity
before the limit in Eq.~(\ref{const}) could start probing $f_a$ at the scales most relevant 
for string theory, $f_a \sim 10^{16}~{\rm GeV}H_{14}$.

Existing constraints on the coupling $f_a$ arise via several known mechanisms. Most notably, 
the CAST experiment directly limits the emission of pseudoscalars from the solar interior 
and has obtained the limit $f_a > 2 \times 10^{10}$ GeV \cite{CAST}, which is quite competitive
with stellar constraints that typically require $f_a \ge 10^{11}$ GeV \cite{Raffelt}.
A massless pseudoscalar-photon coupling in the intergalactic 
magnetic field provides an additional means of probing $f_a$ \cite{conversion} which is,
however, strongly dependent on the assumptions concerning the strength of magnetic field, its
redshift behavior and the number density of free electrons. 
Thus we observe that for $H_{14} \sim O(1)$, the constraint (\ref{const})
derived in this paper is more stringent than any other constraint on massless 
pseudoscalars by at least  two orders of magnitude.

{\it Concluding Remarks}:--- Despite all the evidence for physics beyond the Standard Model 
in the neutrino sector and in cosmology, there is no compelling need for new  
degrees of freedom at or below the electroweak scale. Indeed, neutrino masses and dark matter can arise
from ultra-heavy degrees of freedom, while dark energy is currently consistent with being just a cosmological constant. 
Nonetheless, many models of new high-energy physics do lead to new light degrees of freedom whose mass
may be protected by symmetry, as is the case for the pseudoscalars considered here, and we have shown
 that CMB physics can potentially be a very important probe of such models.
The stochastic fluctuations of a light pseudoscalar field generated during inflation can  
induce the conversion of $E$-mode to $B$-mode polarization in the CMB, and with the current
upper bounds this already leads to stringent constraints (\ref{const}).
It is anticipated that forthcoming experiments such as BICEP will
significantly improve the sensitivity to $H/(2\pi f_a)$ provided the background 
due to lensing can be handled appropriately, and thus these experiments may provide the
primary sensitivity to new physics of this type. 

The search for primordial gravitational waves using  $B$-modes
has intrinsic limitations in that the effect scales as $(H/M_{\rm Pl})^2$, 
and so for $H$ below $10^{13}$ GeV the detection of tensor modes in 
the CMB becomes problematic. In this respect, the search for pseudoscalar fluctuations provides
an independent motivation for studying $B$-modes, which is less predicated on the scale of 
inflation being large. Indeed, with the existing constraints on $f_a$ \cite{CAST}, 
$B$-modes may be generated at an observable level within a wide class of 
inflationary models with $ 10^{10}~ {\rm GeV} \la H \la 10^{14}~ {\rm GeV}$. 
However, since both pseudoscalars and gravitational waves (when combined with the background contribution
from lensing) can have similar power spectra up to normalization, 
it will be important to investigate whether the two can be distinguished 
observationally \cite{prs2}.

\acknowledgments
{\it Acknowledgments}:---We thank N. Afshordi, N. Kaloper, V. Mukhanov 
and S. Speziale for helpful discussions. The work of M.P. and A.R. was supported in part by NSERC of Canada,
and research at the Perimeter Institute is also supported in part by NSERC 
and by the Province of Ontario through MEDT.

\end{document}